\documentclass[fleqn,12pt,twoside]{article}
\usepackage{espcrc1}

\usepackage{graphicx}
\usepackage[figuresright]{rotating}

\newcommand{\AmS}{{\protect\the\textfont2
  A\kern-.1667em\lower.5ex\hbox{M}\kern-.125emS}}

\title{Flow and non-flow event anisotropies at the SPS}

\author{J. Sl\'{\i}vov\'a\address[uni]{University of Heidelberg and MPI f\"ur Kernphysik
Heidelberg, Germany} for the CERES/NA45 Collaboration\footnote{For the full CERES/NA45 Collaboration author list and acknowledgments see the
contribution by J.~P.~Wessels (D. Adamov\'a {\it et al.}) in this volume.}}
\begin{document}
\maketitle

\begin{abstract}
A study of differential elliptic event anisotropies ($v_2$) of 
charged particles and high-$p_T$ pions 
in 158 AGeV/c Pb$\!+\!$Au collisions is presented. Results from  
correlations with respect to the event plane and from two-particle 
azimuthal correlations are compared. The latter give 
systematically higher $v_2$ values at $p_T\!>$1.2~GeV/c 
providing possibly an evidence of a non-flow semihard component. 
\end{abstract}

\section{MOTIVATION} 
\vspace{-0.2cm}
Systematic investigation of elliptic event anisotropies  can shed
light on our knowledge of the equation of state and properties of
nuclear matter under extreme conditions created in ultra-relativistic
heavy-ion collisions. 
Using a Fourier decomposition of the azimuthal distribution of emitted particles 
with respect to the event plane angle ($\Psi_R$) 
the elliptic anisotropy is quantified by the second Fourier coefficient $v_2$ \cite{e877,VolZhang}. 
However an estimate of the event plane, 
{\it a priori} unknown, and dispersion corrections have to be performed \cite{Barette97,PoskVol98}. 
An alternative technique \cite{VolZhang} circumventing these difficulties is 
to extract $v_2$ from the pair-wise azimuthal distribution of the emitted particles 
since the correlation of particles with the event plane induces correlations 
among particles into which $v_2$ enters quadratically. 
Both methods are sensitive to various 'non-flow' effects \cite{DinhOlli}
and one of our aims here is to investigate contributions of semihard processes.

\section{DATA ANALYSIS}
\vspace{-0.2cm}
A large data set of 42$\cdot$10$^6$ $\!$Pb$+$Au collisions 
at 158 AGeV/c was taken by CERES in 1996 at the top 30$\%$ of the geometric cross section. 
$\!\!\!$The experiment with its full azimuthal acceptance is well suited to study azimuthal 
anisotropies of charged particles and  high-$p_T$ pions. Charged particles are reconstructed 
combining the information from two silicon drift detectors (SDD) and a MWPC placed behind the magnetic field. 
Pions with $p\!>$4.5~GeV/c are visible in the RICH detectors ($\gamma_{th}\!\approx\!32$)
and distinguished from electrons by a non-asymptotic ring radius. 
The SDD detectors are used for the event plane measurement by dividing the azimuthal 
acceptance into 100 samples arranged in 4 groups. $\!$The non-uniformities in the  $\Psi_{\!R}$-distribution caused
by dead regions in the detectors, beam and geometrical offsets 
are removed by standard correction procedures \cite{Barette97,PoskVol98}. $\!$The correction factors
for the event plane dispersion go from 3 to 6 depending on centrality.

\section{CORRELATIONS WITH RESPECT TO THE EVENT PLANE}
\begin{figure}[h!]
\vspace{-0.85cm}
\begin{tabular}{lcr}
\hspace{-0.4cm}
\begin{minipage}[t]{50mm}
\includegraphics[height=5.79cm]{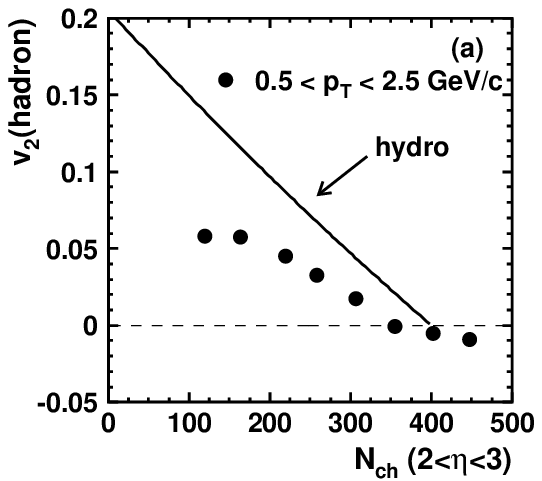}
\end{minipage}
&
\hspace{0.55cm}
\begin{minipage}[t]{50mm}
\includegraphics[height=5.79cm]{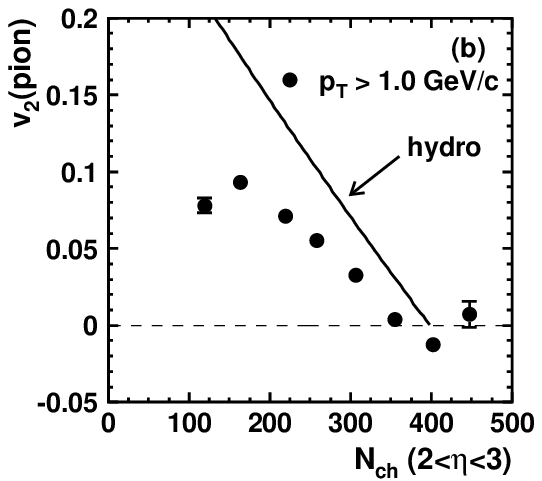}
\end{minipage}
&
\hspace{0.8cm}
\begin{minipage}[t]{35mm}
\vspace*{-6.4cm}
\caption{Centrality dependence of $v_2$ for charged hadrons (a) and pions (b)
together with hydrodynamical calculations \cite{Huovinen}.}
\label{v2nch}
\end{minipage}
\end{tabular}
\end{figure}

\vspace{-0.2cm}
Elliptic anisotropy of charged particles and high-$p_T$ pions
obtained from the event plane analysis has been extensively studied 
as a function of centrality and $p_T$ \cite{thesis}. 
It is found that $v_2$ is decreasing with centrality (Fig.\ref{v2nch}). 
The hydrodynamical calculations \cite{Huovinen} overpredict the data.

\vspace{-0.38cm}

\begin{figure}[h!]
\vspace{-1.8cm}
\begin{tabular}{lcr}
\hspace{-0.4cm}
\begin{minipage}[t]{50mm}
\includegraphics[height=5.79cm]{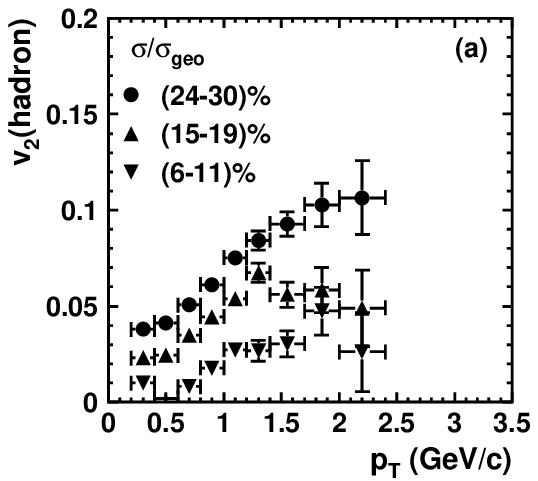}
\end{minipage}
&
\hspace{0.55cm}
\begin{minipage}[t]{50mm}
\includegraphics[height=5.79cm]{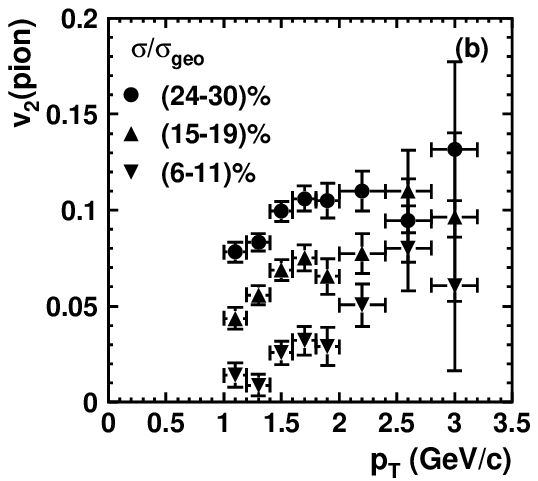}
\end{minipage}
&
\hspace{0.8cm}
\begin{minipage}[t]{36mm}
\vspace*{-6.4cm}
\caption{Dependence of $v_2$ on $p_T$ for charged hadrons (a) and pions (b) for three 
centrality classes.}
\label{v2pt}
\end{minipage}
\end{tabular}
\vspace{-1.0cm}
\end{figure}

\begin{figure}[h!]
\vspace{-0.9cm}
\begin{tabular}{lr}
\begin{minipage}[h]{70mm}
\hspace{-0.3cm}
\includegraphics[height=6.0cm]{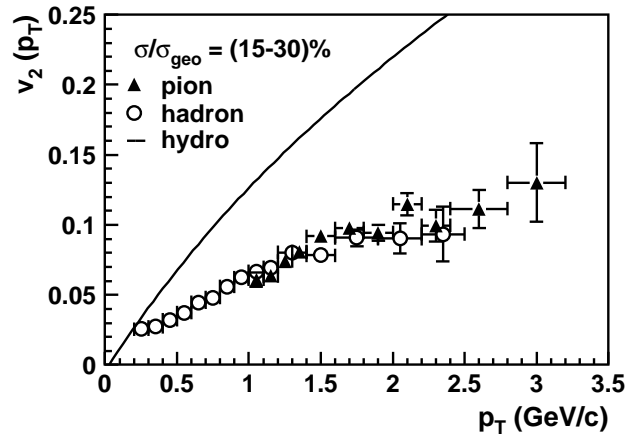}
\vspace*{-1.1cm}
\hspace{1.5cm}
\end{minipage}
&
\hspace{0.6cm}
\begin{minipage}[h]{75mm}
\vspace{-3.8cm}
\caption{Combined data of $v_2(p_T)$ for charged hadrons and identified pions with 
hydrodynamical predictions \cite{Huovinen}. The data were corrected for HBT effects.}
\label{v2-ceres-na49}
\end{minipage}
\end{tabular}
\end{figure}

The transverse momentum dependence of $v_2$ for different centrality classes 
(Fig.\ref{v2pt}) shows a linear increase which levels off at $p_T\approx$~2~GeV/c.
The onset of saturation is clearly visible in the combined data of hadrons and pions 
in Fig.\ref{v2-ceres-na49}, which were corrected for HBT effects \cite{OlliHBT}
calculated  with input from \cite{cereshbt}. The relative corrections 
were found to vary between --15$\%$ at $p_T\!=$~0.25~GeV/c and +10$\%$ at $p_T\!>$1~GeV/c.

\clearpage
\section{TWO-PARTICLE AZIMUTHAL CORRELATIONS}
\begin{figure}[h!]
\vspace{-0.7cm}
\begin{tabular}{lcr}
\hspace{-0.45cm}
\begin{minipage}[h]{50mm}
\includegraphics[height=5.79cm]{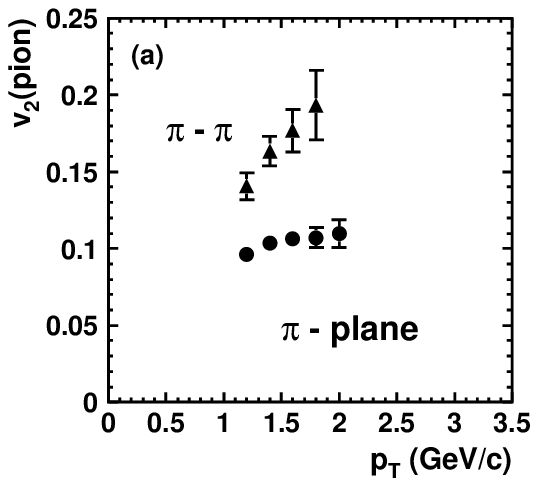}
\end{minipage}
&
\hspace{0.15cm}
\begin{minipage}[h]{50mm}
\includegraphics[height=5.79cm]{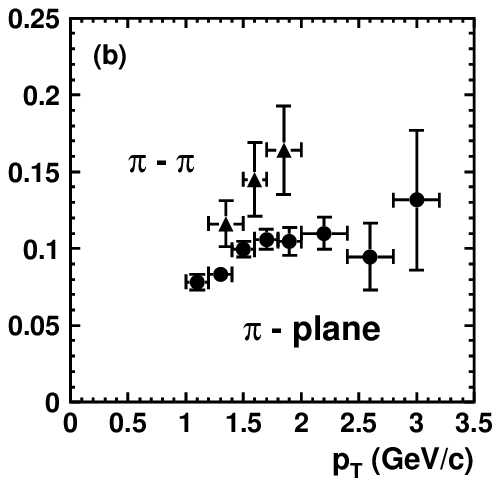}
\end{minipage}
&
\hspace{0.8cm}
\begin{minipage}[h]{40mm}
\vspace{-1.6cm}
\caption{Comparison of $v_2$ from two-particle correlations and 
event plane analysis in centrality (24-30)$\%$ for all $p_T$ above a certain $p_T$-threshold 
given as the abscissa (a) and for differential $p_T$-bins (b).}
\label{v2-rp-cor}
\end{minipage}
\end{tabular}
\vspace{-0.5cm}
\end{figure}

The measured azimuthal correlation of high-$p_T$ pions ($p_T>$1.2~GeV/c) was corrected
for pion detection efficiency and the  finite two-ring 
resolution of the RICH detectors using Monte-Carlo simulations \cite{thesis}. 
The $p_T$-dependence  of $v_2$ parameters obtained 
from the fit (Eq.(36) of \cite{PoskVol98}) is presented in Fig.\ref{v2-rp-cor} 
and compared to the results from the event plane analysis for both threshold (a) and differential
analysis (b).  
The two-particle correlations show systematically higher $v_2$ values indicating the presence
of a non-flow component, and the observed difference grows with $p_T$. 
The ratio $v_2(\pi\!-\!\pi)/v_2(\pi\!-\!\rm{plane})$ in the semicentral collisions
is 1.39$\pm$0.07 averaged over the differential points (Fig.\ref{v2-rp-cor}b), which is
significant compared to a downward correction due to HBT of $\approx$10$\%$ (not applied).

\begin{figure}[h!]
\vspace{-0.8cm}
\begin{tabular}{lr}
\hspace{-0.4cm}
\begin{minipage}[t]{50mm}
\includegraphics[height=6.7cm]{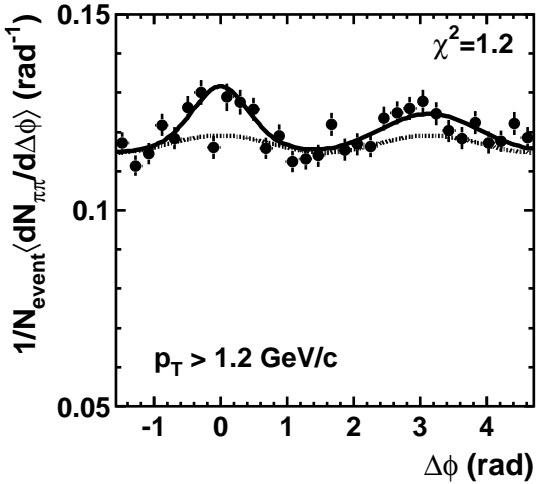}
\end{minipage}
&
\hspace{2.0cm}
\begin{minipage}[t]{85mm}
\vspace*{-7.3cm}
\caption{Two-pion azimuthal correlation for $p_T\!\!>\!\!$~1.2~GeV/c 
in semicentral collisions ($\sigma/\sigma_{geo}$~=~(24-30)$\%$) after 
efficiency correction, where $\Delta\phi$ is the azimuthal angle difference
between pairs of emitted particles.
The full line is a superposition of a background contribution
modulated by elliptic flow (v$_2$~=~(9.0$\pm$0.2)$\%$ from the event plane analysis, dashed line) 
and two gaussians around $\Delta\phi$=0,$\pi$ with amplitudes
and widths as free fit parameters.}
\label{2partcor}
\end{minipage}
\end{tabular}
\vspace{-0.7cm}
\end{figure}

\begin{figure}[h!]
\begin{tabular}{lr}
\begin{minipage}[t]{50mm}
\includegraphics[height=5.8cm]{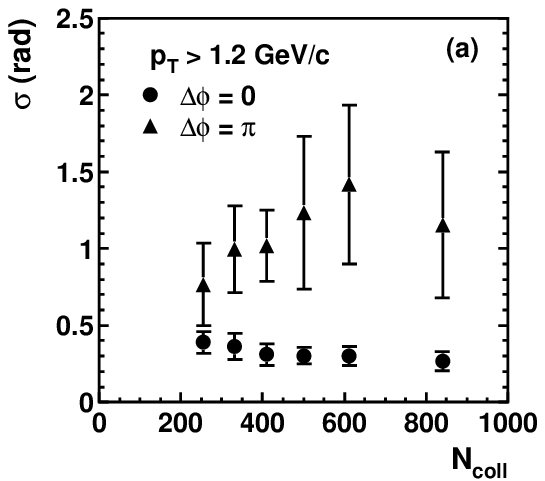}
\end{minipage}
&
\hspace{1.5cm}
\begin{minipage}[t]{50mm}
\includegraphics[height=5.8cm]{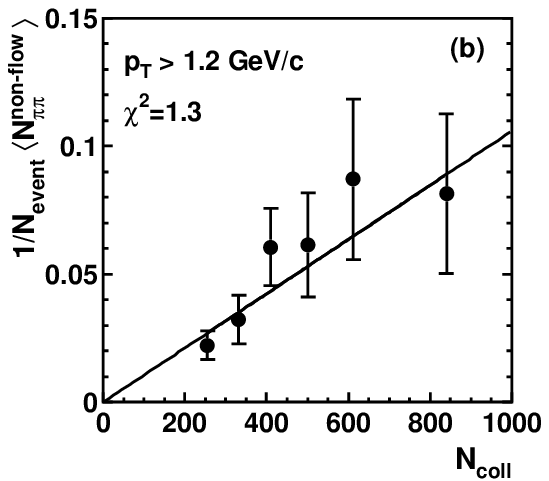}
\end{minipage}
\end{tabular}
\vspace{-0.7cm}
\caption{Centrality dependence of the gaussian width of the correlation peaks 
at $\Delta\phi$=~0 and $\Delta\phi$=$\pi$ (a) and of the non-flow yield of pion pairs, 
which is the sum of the areas under the gaussian peaks calculated from the fit parameters (b).}
\label{yield-broadening}
\vspace{-0.5cm}
\end{figure}

Assuming the non-flow contribution is due to an additional physics process (e.g. resonance decays,
semihard scattering) the data were fitted by two gaussians at $\Delta\phi$=0, $\pi$ sitting on top
of the flow-modulated background (Fig.\ref{2partcor}).
The widths of the gaussian peaks and their centrality dependence are different. 
The 'back-to-back' peak is broader than the 'near-angle' peak and its width 
increases with centrality (Fig.\ref{yield-broadening}a) 
whereas the near-angle peak retains its width. The 'non-flow' yield 
of pion pairs (Fig.\ref{yield-broadening}b), which is the sum of the 
areas under the gaussian peaks calculated from the fit parameters, grows linearly with number 
of binary collisions ($N_{coll}$). 

\clearpage

An interpretation of the non-flow component in terms of resonance decays 
seems rather unlikely in view of the high invariant mass required
($\approx\!$~2.5~GeV/c$^2$). An explanation by minijet production 
\cite{Accardi} is suggested by $\sigma(\Delta\phi\!=\!\pi)\!\propto\!N_{coll}$
indicative of semihard rescatterings, $\sigma(\Delta\phi\!=\!0)\!\approx\!const.$ as expected
for fragmentation, and the scaling of the absolute $\pi\!-\!\pi$ yield
with $N_{coll}$.

\section{CONCLUSIONS}
The presented results on elliptic event anisotropies of charged particles 
and high-$p_T$ pions cover a wide range of $p_T$ and show a saturation behaviour
around $p_T\approx$~2~GeV/c. 
The observed non-flow component grows with $p_T$ and is significantly larger than potential
contribution from the HBT correlations.
The data suggest an explanation based 
on semihard processes.

\end{document}